\documentclass{ws-ijmpc}
\usepackage[super]{cite}
\usepackage{xcolor}
\usepackage{subfigure}
\usepackage[verbose,hypertexnames=false]{hyperref}
\hypersetup{colorlinks=false,allbordercolors=blue,pdfborderstyle={/S/U/W 1}}

\begin{document}

\markboth{XF Jiang et al.}
{Effects of analyst sentiment on volatility dynamics in financial market}

\catchline{}{}{}{}{}

\title{Effects of analyst sentiment on volatility dynamics in financial market}

\author{Xiongfei Jiang\footnote{
Corresponding author.}}
\address{
College of Finance, Ningbo University of Finance \& Economics\\ Ningbo, 315175, PR China, jiangxiongfei@nbufe.edu.cn}

\author{Tao Cen}
\address{Institute of Digital Economy and Public Policy, Ningbo University of Finance \& Economics\\
Ningbo, 315175, PR China,
centao@nbufe.edu.cn}

\author{Ling Bai}
\address{
	College of Finance, Ningbo University of Finance \& Economics\\ Ningbo, 315175, PR China,	
	bailing@nbufe.edu.cn}

\author{Lifu Jin}
\address{School of Economics and Management, Wenzhou University of Technology\\
	Wenzhou, 325035, PR China,
	jinlifu@zju.edu.cn}

\author{Jiu Zhang}
\address{School of Digital Economy and Management, Fuyao University of Science and Technology\\
	Fuzhou, 350109, PR China,
	zhangjiu@zju.edu.cn}

\author{Long Xiong\footnote{
		Corresponding author.}}
\address{School of Physics and Astronomy, Yunnan University\\
	 Kunming, 650091, PR China,
	xionglong@zju.edu.cn}

\maketitle


\begin{abstract}
Text emotions are extracted using natural language processing technique on a substantial corpus of analyst reports on the Chinese stock market. Subsequently, the text-based analyst sentiment indices are constructed. It is observed that both optimistic and pessimistic sentiments represent short-range memory. Optimistic and pessimistic sentiments are correlated with volatility positively and negatively, respectively.
The analysis of transfer entropy reveals that past pessimistic sentiment affects future volatility.
Further, we model the driving effect of analyst sentiment on volatility using a GARCH model. The results show that pessimistic sentiment is an explanatory factor for volatility, while optimistic sentiment is not.
\end{abstract}

\keywords{financial market; analyst sentiment; volatility; transfer entropy; GARCH}

\ccode{PACS number: 89.65.Gh}

\section{Introduction}


As a core information carrier that assists investors in evaluating the fundamentals of underlying stocks in financial markets, analyst reports have long served as an indispensable reference for investment decision-making. Over the past decades, analysts have disclosed multi-dimensional information through research reports, including quantifiable indicators such as earnings forecasts, buy-hold-sell investment ratings, as well as qualitative interpretations of corporate development logic and industry trends \cite{ram08}. Existing studies on analyst reports mostly focus on the structured data dimension within reports, which has been proven to exert a significant impact on the dynamic operation of financial markets \cite{wom96, giv79, lys90, bra03, asq05, giv09, cal13, ert03}.


Although the textual data in analyst reports has presented apparent significance in finance already, it has been largely overlooked in the extant literature \cite{ram08}. 
At present, information contained in textual data can be quantified through natural language processing.
For instance, textual opinions extracted from analyst reports are correlated with the performance of stock markets \cite{hua14}. Moreover, there is a decline in analyst sentiment as the fiscal month progresses, which in turn affects the accuracy of earning forecasts \cite{jia22}.
In addition, it is demonstrated that other textual data, such as earnings press releases \cite{dav12}, annual reports \cite{fel10, lou11}, initiation reports \cite{twe12}, and conference reports \cite{bus11}, contain measurable information for forecasting of stock price and fundamentals. 
It is, therefore, feasible to extract meaningful information from analyst reports that reflect investor sentiment in financial markets.

The proxies of investor sentiment can be classified into two categories: objective and subjective. Subjective proxies quantify the psychological states of investors directly, primarily through questionnaires and interviews \cite{lem06,qiu04}.
Nevertheless, the survey sentiment cannot predict the price well due to the delay of questionnaire surveys \cite{fis00}. 
In contrast to subjective proxies, objective proxies of investor sentiment do not directly measure investor sentiment. Instead, they refer to a variety of real-time data that describe market states. The most commonly used objective proxies include mutual fund flows \cite{bro03}, trading volume \cite{bak04}, dividend premiums \cite{bak04a, bak04b}, closed-end fund discounts \cite{nea98}, and option implied volatility \cite{wha00}. In order to comprehensively reflect investor sentiment, the principal component analysis is introduced to construct the composite indicator \cite{bak06}.
As one of the most important information providers, analyst sentiment will be a beneficial proxy to
improve the market sentiment. Thus, it is necessary to accurately measure analyst sentiment in real-time, i.e., extracting it from analyst reports.

Behavioral finance deems that stock markets are not necessarily efficient. Noise traders make trading decisions with psychological biases, significantly affecting price returns and volatility \cite{bla86,bak06,chu12,de90,meh02}.
The investor sentiment may act as a systematic risk factor in the pricing process \cite{lee02}. Rationally uninformed traders sometimes amplify the impact of sentiment on markets and drive asset prices to deviate from the underlying value \cite{men12}.
From a general perspective, the effect of exogenous information on stock markets are investigated with various emerging data recently \cite{che17,che18,jia13,zha24,jin22}.
Therefore, measuring the effect of analyst sentiment on financial dynamics is vital for both financial theory and industry.

In this paper, we extract sentiment from analyst reports in the Chinese stock market with natural language processing,
and compile the analyst sentiment indices on a weekly time scale. Both optimistic and pessimistic sentiments display a short-range auto-correlation in time. Pessimistic sentiment is negatively correlated with volatility.
With the transfer entropy, it is observed that past pessimistic sentiment affects future volatility.
Further analysis shows that pessimistic makes a significant informative contribution to the volatility process in a GARCH framework.

\section{Data and method}

We collect abstracts of analyst reports from Sina Finance from April 2013 to December 2015 in the Chinese stock market. 
In order to assess the predictive ability of the reports, only those published prior to the release of the annual report are taken into consideration.
Meanwhile, to avoid the arbitrariness of reports, we select those widely watched stocks to ensure the reliability of results. Consequently, stocks are excluded if the number of reports is less than $20$ in each fiscal year. The final sample consists of $7789$ reports covering $197$ stocks.


For the task of emotional tendency identification across large-scale unstructured text corpus, natural language processing technology has provided a mature, computer-driven solution that avoids the inefficiency of traditional manual annotation. In this study, we adopt the open-source Jieba text segmentation toolkit (available at https://github.com/fxsjy/jieba) as the core text processing component. With its proven reliable performance in extracting optimistic and pessimistic sentiment orientations from short-form texts, the toolkit is fully applicable to quantifying the overall emotional attributes embedded in full-text analyst reports. We further introduce a manually curated Chinese sentiment vocabulary base, paired with a customized set of linguistic rules targeting Chinese expression patterns, to accurately identify complex emotional phrases including negative-modified terms (e.g. "not optimistic") and degree-adverb intensified terms (e.g. "very pessimistic") \cite{zha10}.
Individual sentiment words are scored from $-2$ to $2$, i.e., $-2$ for very pessimistic, $-1$ for pessimistic, $0$ for neutral, $1$ for optimistic, and $2$ for very optimistic. The optimistic and pessimistic sentiment counts in a single report are summed up as the report's optimistic and pessimistic sentiment values, respectively. Then we construct the weekly time-scale analyst sentiment indices
\begin{equation}\label{sentimentIndex}
	S^{+}\left( t\right)=\frac{1}{\mathcal{N}_t} \sum_{i}\sum_{m\in t} s_i^{+}\left(m\right),
\end{equation}
and
\begin{equation}\label{sentimentIndex}
	S^{-}\left( t\right)=\frac{1}{\mathcal{N}_t} \sum_{i}\sum_{m\in t} | s_i^{-} \left(m\right)|,
\end{equation}
where $s_i^{\pm} \left(m\right)$ are the values of optimistic and pessimistic sentiments of the $m$-th reports on the $i$-th stock, and $\mathcal{N}_t$ is the total number of reports released in the $t$-th week.
It should be noted that both values of optimistic and pessimistic sentiments are positive.

\begin{figure}[!hbt]
	\centering
	\includegraphics[height=0.7\textwidth,clip]{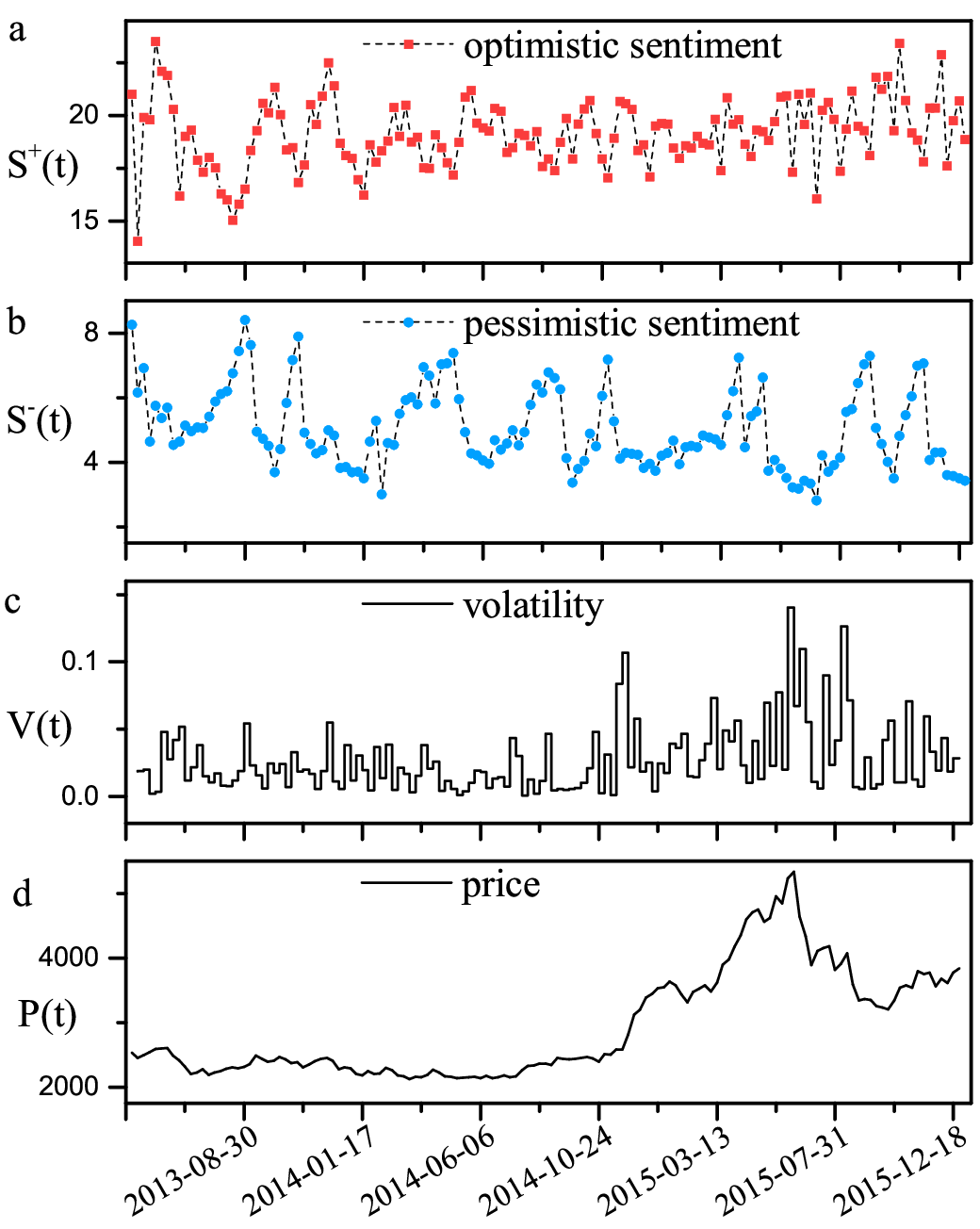}
	\protect\caption{Demonstration of variables.
		(a) and (b) are the optimistic and pessimistic analyst sentiment indices, respectively. (c) and (d) show volatility and price for CSI 300 Index, respectively.  }
	\label{seriesALL}
\end{figure}

Meanwhile, we collect the weekly closed price of several typical indices in the Chinese stock market during the corresponding period, including the CSI 300 Index, SSE 50 Index, and CSI Smallcap 500 Index. These represent composite stocks, blue-chip stocks, and smallcap stocks, respectively. Denoting the financial index at time $t$ as $P \left( t \right) $, return and volatility are defined as
$R(t)=lnP(t+1)-lnP(t)$ and
$V \left( t\right) = \left| R\left( t\right) \right| $, respectively.
The variables of analyst sentiment, prices, and volatility of CSI 300 are demonstrated in Fig.~\ref{seriesALL}.
The descriptive statistics of analyst sentiment and all stock indices are listed in Table~\ref{statistics}.

\begin{table}[h]
	\centering
	\begin{tabular}{lllllllll}
		\hline
		Variable              & Mean    & SD    & Min    & Median & Max   & Skewness & Kurtosis \\
		\hline
		Optimistic sentiment  & 19.13   & 1.625 & 14.02  & 19.15  & 23.51 & -0.051 & 3.409 \\
		Pessimistic sentiment & 5.039   & 1.235 & 2.809  & 4.711  & 8.403 & 0.623  & 2.601 \\
		Return of CSI 300     & 0.003   & 0.037 & -0.140 & 0.006  & 0.107 & -0.716 & 5.352 \\
		Return of SSE 50      & 0.002   & 0.040 & -0.143 & 0.001  & 0.139 & 0.001  & 5.171 \\
		Return of Smallcap    & 0.007   & 0.044 & -0.180 & 0.013  & 0.131 & -1.082 & 6.074\\
		\hline
	\end{tabular}
	\caption{Descriptions and Statistics of data. All values are calculated for weekly intervals. 
		Each variable contains $141$ datapoints. }  \label{statistics}
\end{table}

\section{Correlation between analyst sentiment and volatility}

\begin{figure}[!hbt]
	\centering
	\includegraphics[height=0.4\textwidth,clip]{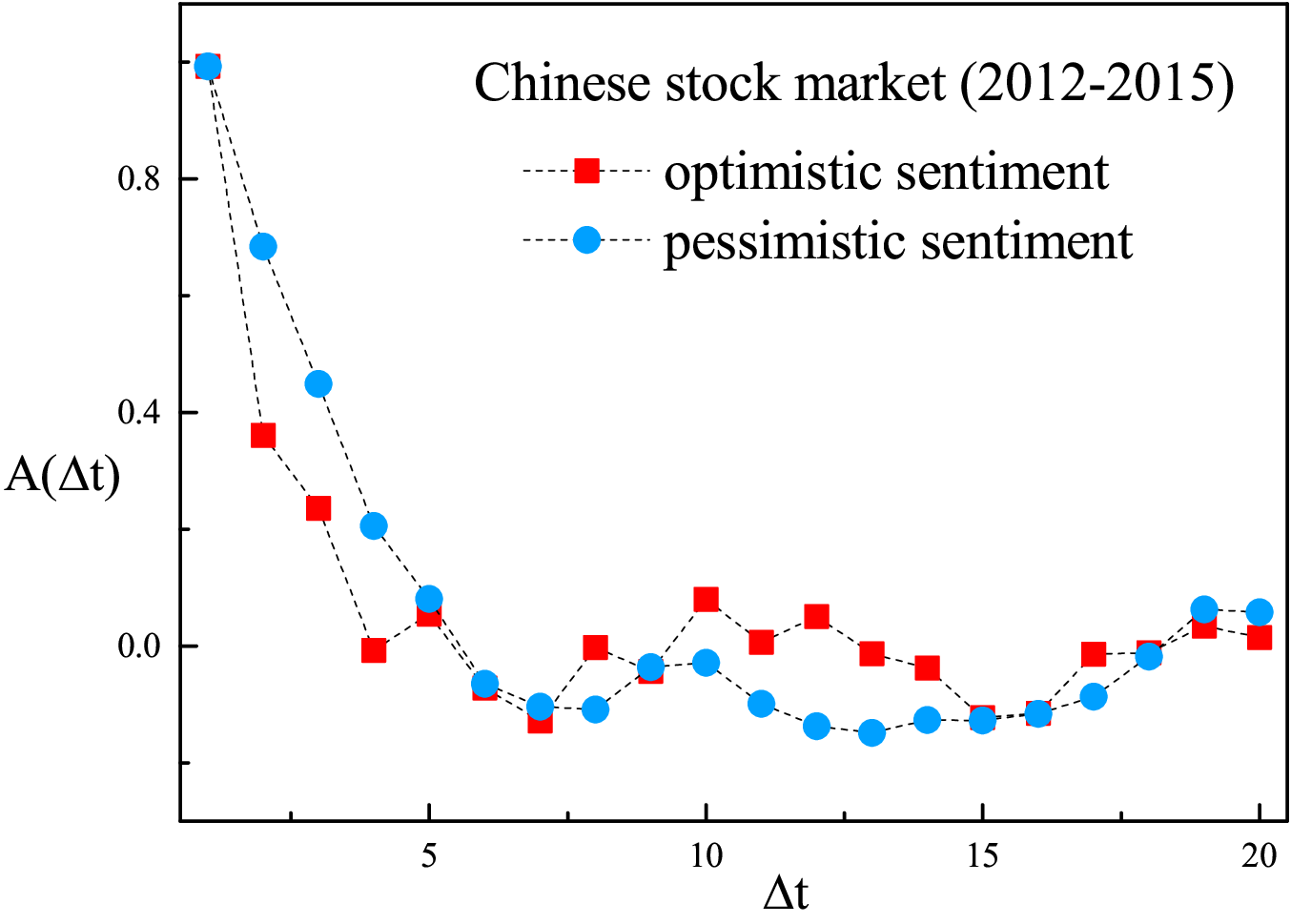}
	\protect\caption{Autocorrelation of optimistic and pessimistic analyst sentiment indices. The correlated times last about $3$ and $4$ weeks for optimistic and pessimistic index, respectively. }
	\label{autocor}
\end{figure}

The auto-correlation function of the analyst sentiment indices $S^{\pm}\left( t \right) $ is defined as
\begin{equation}\label{autocorr}
A\left( \Delta t \right) = \frac{\left\langle S\left( t \right) S\left( t + \Delta t \right)\right\rangle -\left\langle S\left( t\right) \right\rangle ^2  }{\sigma}
\end{equation}
where $ \left\langle \cdots \right\rangle $ is the average over time $t$, and $\sigma$ denotes the standard deviation of the time series. It is well known that volatility in financial dynamics is long-range correlated in time, i.e., it decays in a power law.
After analyzing the auto-correlation function of analyst sentiments, we observe that the curves both for optimistic and pessimistic sentiments show a short-range correlation.
As shown in Fig. \ref{autocor}, the auto-correlation of pessimistic sentiment could last for $5$ weeks, while the one of optimistic sentiment lasts up to $3$ weeks.
Thus, the auto-correlation of pessimistic sentiment is relatively stronger than that of optimistic sentiment.

To quantify the dependency relation between sentiment and volatility preliminarily,
we define the sentiment-volatility correlation function as
\begin{equation}\label{svcorr}
G\left( \Delta t \right) = \frac{\left\langle S\left( t \right) V\left( t + \Delta t \right)\right\rangle -\left\langle S\left( t\right) \right\rangle \left\langle V\left( t\right) \right\rangle  }{\sigma_s \sigma_v},
\end{equation}
where $ \left\langle \cdots \right\rangle $ is the average over time $t$ and $\sigma$ denotes the standard deviation of the time series. $G\left( \Delta t \right) > 0$ indicates that analyst sentiment has a positive driving effect on volatility, while $G\left( \Delta t \right) < 0$ suggests that analyst sentiment has a negative effect on volatility. 

\begin{figure}[!hbt]
	\centering
	\subfigure{\label{a}\includegraphics[width=0.495\textwidth,clip]{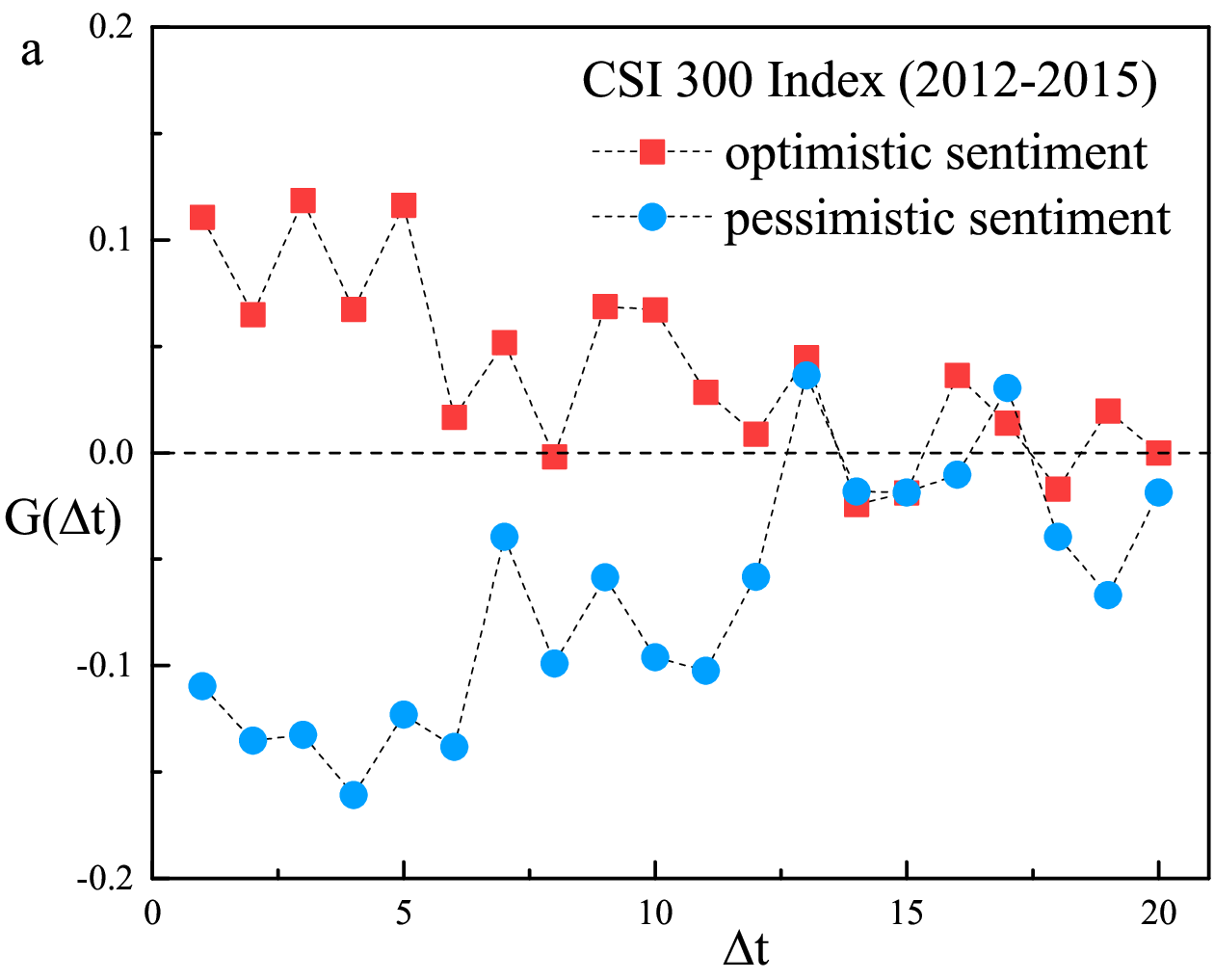}}
	\subfigure{\label{b}\includegraphics[width=0.495\textwidth,clip]{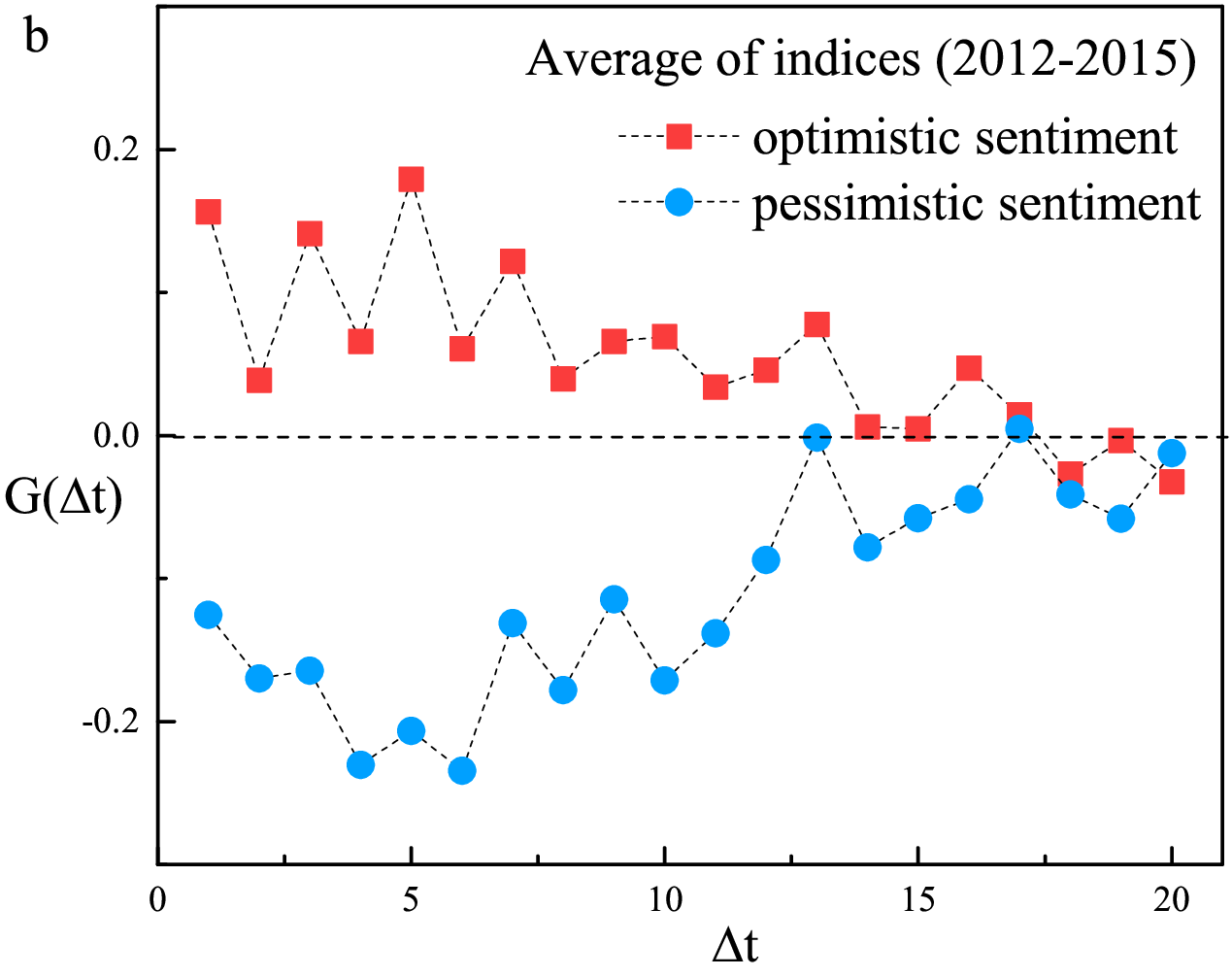}}
	\protect\caption{(a) Sentiment-volatility correlation functions for the CSI 300 Index. (b) The average sentiment-volatility correlation functions over indices in the Chinese market. Optimistict sentiment is positively correlated with volatility, while the pessimistic one is negatively correlated with volatility. }
	\label{senti_volatility_cor}
\end{figure}

As shown in Fig. \ref{senti_volatility_cor}a, 
pessimistic sentiment is negatively correlated with volatility for the CSI 300 Index. It means that a high pessimistic sentiment decreases future volatility, while a low pessimistic sentiment increases future volatility.
Conversely, optimistic sentiment is positively correlated with volatility. It suggests that high optimistic sentiment induces higher volatility, while low optimistic sentiment may lead to low volatility.
Meanwhile, the correlation strength for pessimistic sentiment is stronger than that for optimistic sentiment.
An interesting phenomenon is that the sentiment-volatility correlation does not fluctuate around zero only when $\Delta t>0$, and is about zero when $\Delta t<0$. It suggests that analyst sentiments affect volatility, but not vice versa. 
Since the behaviors of the other indices in the Chinese market look essentially the
same, an average over indices has been taken in Fig.\ref{senti_volatility_cor} b.

What should be the social origin of the negative (positive) correlation between pessimistic (optimistic) sentiment and volatility?
An explanation is investors' bounded rationality and risk aversion in financial markets \cite{kah03}. Investors rush for trading that increases volatility \cite{qiu06,jia17a}, as market sentiment increases (i.e., high optimistic or low pessimistic sentiment ). When market sentiment drops (i.e., low optimistic or high pessimistic sentiment ), investors stay inactive in trading that reduces volatility \cite{qiu06,jia17a}.

In addition, we investigate the correlations between analyst sentiment and other market variables, including return, volume, and turnover. Analyst sentiment is not significantly correlated with these variables. Although some studies show that investor sentiment affects daily returns \cite{bak06,chu12}, the dissenting voice argues that it is unrelated to returns at the weekly and monthly time scale \cite{bro04}.
Given that the data in this paper are on a weekly time scale, the results are consistent with ref. \cite{bro04}. 


To accurately specify the information flow between two variables, it is necessary to introduce a rigorous mathematical term to 
replace the simple linear correlation. In this case, the transfer entropy is a good proxy, since the financial time series represents non-stationary characteristics.
Considering two processes $X$ and $Y$, transfer entropy from $Y$ to $X$ is defined as follows:
\begin{equation}\label{TE}
T_{Y \to X}=\sum p\left(X_{t+1},X_{t}^k,Y_{t}^l \right)log\frac{p\left(X_{t+1}|X_{t}^k,Y_{t}^l \right) }{p\left(X_{t+1}|X_{t}^k \right)},
\end{equation}	
where $X_{t}$ and $Y_{t}$ are the states at time $t$ of variables $X$ and $Y$, respectively. $p\left(X,Y \right)$ is the joint probability of $X$ and $Y$, and 
\begin{equation}\label{TE1}
p\left(X_{t+1},X_{t}^k,Y_{t}^l \right)=p
\left(X_{t+1},X_{t},\cdots, X_{t-k+1}, Y_{t}, Y_{t-l+1} \right) 
\end{equation}
is the joint probability distribution of state
$X_{t+1}$ with its $k+1$  predecessors, and with the $l$ predecessors of state $Y_{t}$.
The definition of transfer entropy assumes that events on a certain week may be influenced by events
of $k$ and $l$ previous weeks. In practice, we take $k=3$ and $l=3$.
Then, we define the net information flow to measure the disparity in influences of two variables as
\begin{equation}\label{netflow}
\delta_{XY}=T_{X\to Y}-T_{Y\to X}.
\end{equation}
If the value of $\delta_{XY}$ is positive, we say that past $X$ influences future $Y$.

We examine the net information flow between analyst sentiment and volatility, $\delta_{SV}$, where $S$ and $V$ denote analyst sentiment and volatility, respectively. As displayed in Fig.~\ref{TE}a, $\delta_{SV}>0$ for the SSE50 Index and pessimistic sentiment, i.e., the information flows from pessimistic sentiment to volatility. In other words, past pessimistic sentiment affects future volatility.
It is verified again that pessimistic sentiment is related to volatility in a unilateral way.
On the other hand, the net information flow between optimistic sentiment and volatility is almost zero. Thus, there is no significant directionality between optimistic sentiment and volatility.
The results also hold for other indices of the Chinese stock market, such as the CSI 300 Index and the Smallcap Index. The average of net information flows over indices $\overline{\delta_{SV}} $ is shown in Fig.\ref{TE}~b, which are quite robust.

\begin{figure}[!h]
	\centering
	\subfigure{\label{a}\includegraphics[width=0.49\textwidth,clip]{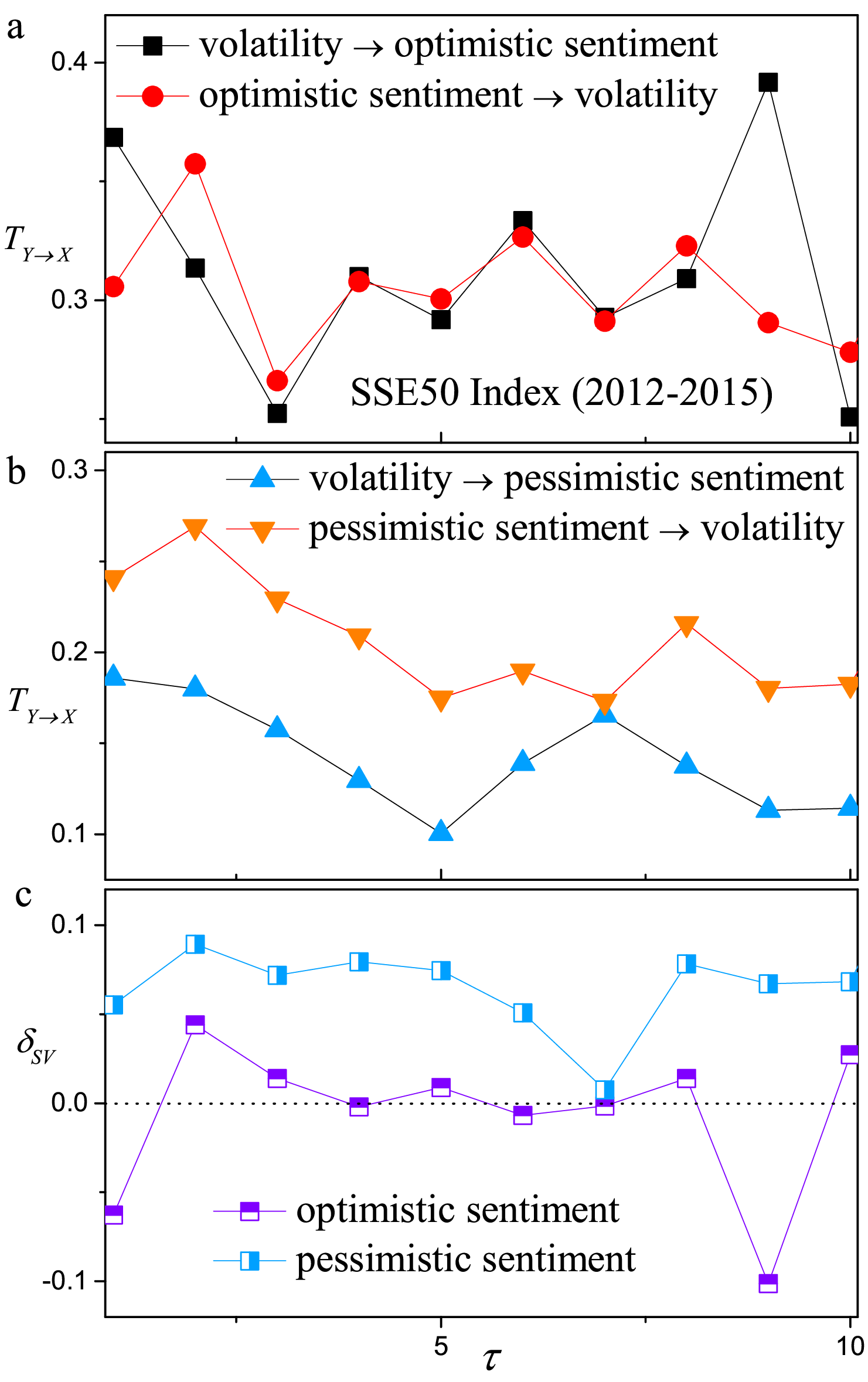}}
	\subfigure{\label{b}\includegraphics[width=0.49\textwidth,clip]{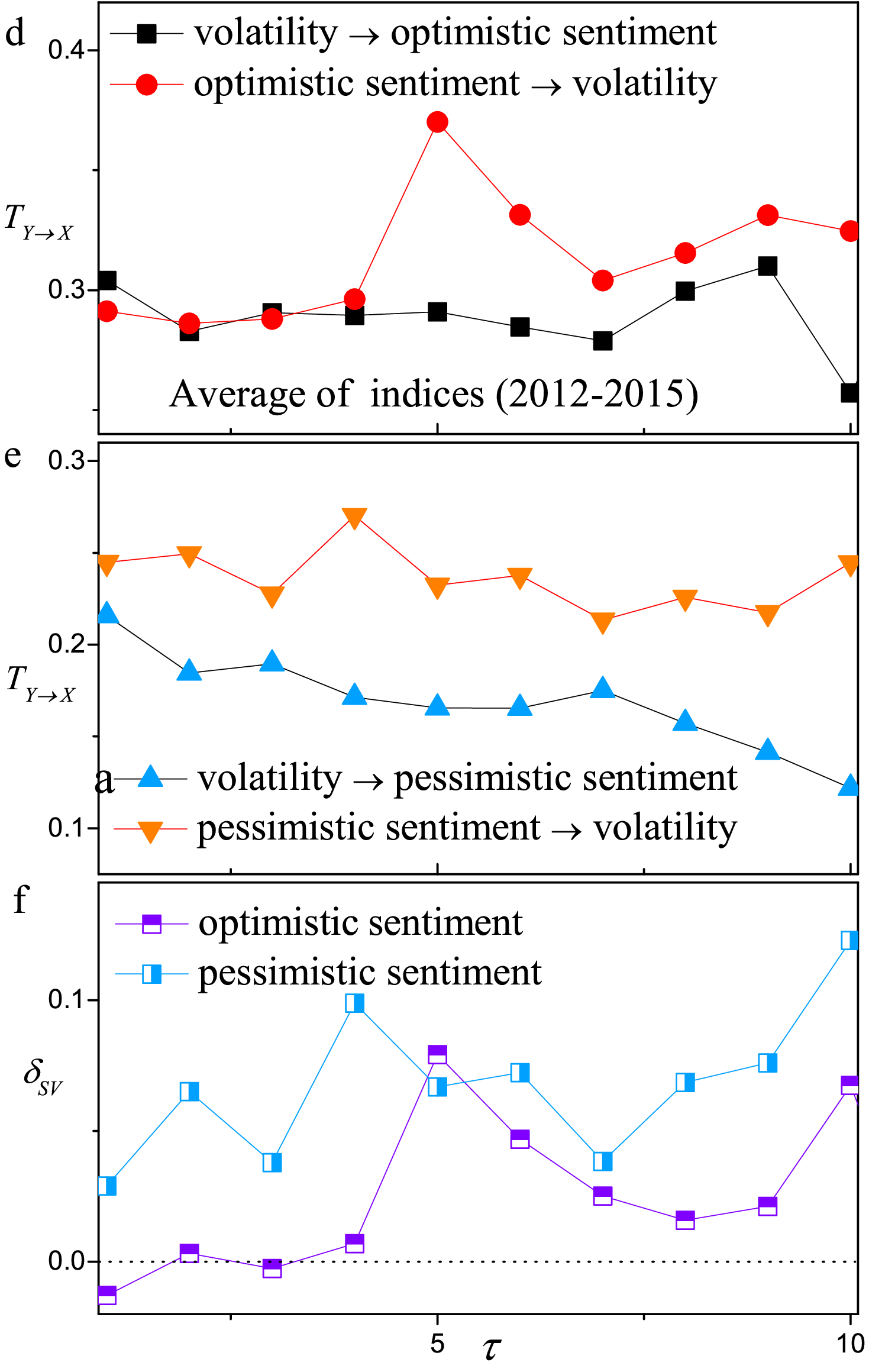}}
	\caption{(a) (b) (c) Transfer entropy and net information flow for the SSE 50 Index. (d) (e) (f) Average transfer entropy and average net information flows over indices in Chinese market.  
	}
	\label{TE}
\end{figure}

Briefly speaking, pessimistic sentiment influences volatility, but not vice versa. 
From this perspective, pessimistic sentiment provides independent information that contributes to the disclosure of information in the market, and is a new proxy of market sentiment.
However, there is merely net information flow between optimistic sentiment and volatility.


\section{GARCH model incorporated with analyst sentiment}

To precisely describe the driving effect of analyst sentiment on volatility, we adopt the GARCH model to deal with the volatility persistence \cite{bol86}. For the specification, the model is restricted to the GARCH $\left( 1,1 \right)$, which is a parsimonious representation of conditional variance and quantifies financial time series quite well \cite{bol86}.
The benchmark model is described by the following:
\begin{equation}\label{garch1}
R\left( t\right)=\mu+\epsilon \left( t\right), \epsilon \left( t\right) | \Omega ~\left( 0, h^2 \left( t\right)\right),
\end{equation}
\begin{equation}\label{garch2}
h^2 \left( t\right)= \omega+\alpha\epsilon^2 \left( t-1\right)+\beta h^2 \left( t-1\right),
\end{equation}
where $R \left( t\right)$ is the return at time $t$, $\mu$ is a constant, $\epsilon\left( t\right)$ is the uncorrelated errors and $h^2 \left( t\right)$ represents the conditional variance of $\epsilon \left( t\right) $. The sum of the coefficients $\alpha + \beta$ indicates the degree of volatility persistence.


The unit root is tested by Dickey and Fuller (DF) and Phillips and Perron (PP). Table \ref{unitroot} shows that all variables pass the test.
As displayed in Table \ref{GrachSpecification}, all benchmark models fit the GARCH $\left( 1,1 \right)$ quite well with statistical significance at the 1\% level. The fact that $\alpha + \beta$ is fairly close to $1$ indicates the persistence of past volatility in explaining current volatility \cite{eng86}.

\begin{table}[h]
	\centering
	\begin{tabular}{llll}
		\hline
		Variable & ADF test  & PP test & Result \\
		\hline
		Optimistic sentiment     & $-8.441^{***}$  & $-8.525^{***}$    & stationary \\
		Pessimistic sentiment    & $-5.332^{***}$  & $-5.548^{***}$    & stationary \\
		Return of CSI 300        & $-9.953^{***}$  & $ -10.010^{***}$  & stationary \\
		Return of SSE 50         & $-10.760^{***}$ & $ -10.820^{***}$  & stationary \\
		Return of Smallcap       & $-9.055^{***}$  & $-9.048^{***}$    & stationary \\
		\hline
	\end{tabular}
	\caption{Unit root of using Dickey and Fuller (DF) test and Phillips and Perron (PP) test. All unit root tests are executed with a drift and time trend. ***, **, * denote a test statistic is statistically significant at the 1\%, 5\%, or 10\% level of significance respectively. }  \label{unitroot}
\end{table}

Then the analyst sentiment indices
$S\left( t\right)$ is incorporated into Eq.(\ref{garch2}) as an exogenous variable, written as
\begin{equation}\label{garch3}
h^2 \left( t\right)= \omega+\alpha\epsilon^2\left( t-1\right)+\beta h^2 \left( t-1\right)+\lambda S\left( t-1\right) .
\end{equation}
Table \ref{GarchSpecification} shows the parameter estimates and diagnostic statistics of the GARCH $\left( 1,1 \right)$ specification with the optimistic and pessimistic sentiment indices, respectively.

\begin{table}[h]
	\centering
	\begin{tabular}{lllll}
		\hline
		Specification              & $\alpha$      & $\beta$       & $\lambda$       &$\alpha+\beta$  \\
		\hline
		CSI 300 Index $$$$ \\
		Benchmark                  &$0.153^{***}$  &$0.810^{***}$  & $N/A$           & $0.963$       \\
		Optimistic sentiment       &$0.151^{**}$   &$0.791^{***}$  & $0.346$         & $0.942$        \\
		Pessimistic sentiment      &$0.070^{*}$    &$0.852^{***}$  & $-1.159^{**}$   & $0.922$        \\
		\hline
		SSE 50 Index $$$$ \\
		Benchmark                  &$0.192^{***}$  &$0.761^{***}$  & $N/A$           & $0.953$       \\
		Optimistic sentiment       &$0.184^{***}$  &$0.762^{***}$  & $0.156$         & $0.946$        \\
		Pessimistic sentiment      &$0.141^{**}$   &$0.803^{***}$  & $-0.475$        & $0.944$        \\
		\hline
		Smallcap Index $$$$ \\
		Benchmark                  &$0.212^{**}$   &$0.748^{***}$  & $N/A$            & $0.960$       \\
		Optimistic sentiment       &$0.208^{**}$   &$0.745^{***}$  & $0.227$         & $0.953$        \\
		Pessimistic sentiment      &$0.204^{**}$   &$0.755^{***}$  & $-0.185$        & $0.959$        \\
		\hline
	\end{tabular}
	\caption{Estimates of the GARCH $\left( 1,1 \right)$ model.
		***, **, * denote a test statistic is statistically significant at the 1\%, 5\%, or 10\% level of significance respectively. } \label{GarchSpecification}
\end{table}

For the CSI 300 Index,
with the incorporation of pessimistic sentiment, the value of $\alpha+\beta$ decreases. It suggests that pessimistic sentiment is a good proxy to forecast contemporaneous volatility. Meanwhile, there is a negative relationship between pessimistic sentiment and volatility, since the value of $\lambda$ is negative.
This result is consistent with the sentiment-volatility correlation.
Specifically, high pessimistic sentiment reduces volatility in the future, while low pessimistic sentiment increases volatility.
By contrast, optimistic sentiment is not statistically significant as an exogenous variable
in the GARCH model.
In brief, after accounting for persistency-in-volatility under the GARCH framework, pessimistic sentiment contributes
significant information on the volatility process of the CSI 300 Index.

The results of the GARCH model for the Smallcap Index and SSE 50 Index also indicate a negative correlation between pessimistic sentiment and volatility, as shown in Table \ref{GrachSpecification}. This finding is consistent with the results observed for the CSI 300 Index. However, pessimistic sentiment is not statistically significant for the Smallcap and SSE 50 Index. It is due to the fact that the sample stocks included in this paper primarily comprise those listed on the CSI 300 Index, and thus, the results may not be representative of other indices.
Consequently, although pessimistic sentiment is negatively correlated with volatility of all indices, it only has predictive power for volatility of the CSI 300 Index.


There are two interesting observations. The first one is that pessimistic sentiment is a significant exogenous variable of the volatility forecasting, while optimistic sentiment is not.
The asymmetric behavior of temporal variation is observed between optimistic and pessimistic sentiments.
Specifically, optimistic sentiment has a prominent bias, and decreases with the fiscal month \cite{jia22}. On the contrary, optimistic sentiment affects the accuracy of analyst earning forecasts only at the end of the fiscal year, but does not relate to the earning forecasts in other fiscal months.
Thus optimistic sentiment is poorly used to predict volatility. In contrast, pessimistic sentiment is more objective and does not vary with the fiscal month \cite{jia22}, and therefore has better predictive power.

The second observation is that pessimistic sentiment negatively affects volatility. 
Beyond the social origin, we try to explain it from the perspective of information disclosure in financial markets.
Analysts affiliated with investment banks are cautious in releasing pessimistic reports and tend to delay them as long as possible, since investment banks are often tied to listed companies \cite{bri05}. For example, analysts usually release reports after the listed companies announce the news, rather than in advance. Before this, there has already been a great deal of inside information or rumors that caused market turmoil. Once the actual information is released, speculation stops in the market.
This will increase confidence and certainty about future expectations and, therefore, reduce future volatility.

\section{Conclusion}

With the analyst reports from 2013 to 2015 in the Chinese stock market, sentiment is extracted from the text with natural language processing.
We compile the weekly optimistic and pessimistic analyst sentiment indices, both of which display a short-range auto-correlation in time.
Pessimistic sentiment is negatively correlated with volatility, while optimistic sentiment is positively correlated.
With the transfer entropy, it is observed that past pessimistic sentiment affects future volatility.
Further analysis shows that pessimistic sentiment makes an important informative contribution to the volatility process in the GARCH framework, while optimistic sentiment can not. This suggests that pessimistic sentiment has predictive power for the volatility dynamics.

\section*{Acknowledgments}
This work was supported by Philosophy and Social Sciences Planning Project of Zhejiang Province under Grant No. 25NDJC027YBM and 21NDQN291YB, National Natural Science Foundation of China under Grants No. 12305053,
the Yunnan Fundamental Research Project under Grant No. 202401CF070167, Fuyao University of Science and Technology Talent Recruitment and Development Program No.PF2025-E07.

\section*{ORCID}
\noindent Xiongfei Jiang - \url{https://orcid.org/0000-0001-5879-6409}

\noindent Tao Cen - \url{https://orcid.org/0000-0002-5673-3495}

\noindent Long Xiong - \url{https://orcid.org/0000-0002-8702-1976}


\begin{thebibliography}{0}




\bibitem{ram08}
S.~Ramnath, S.~Rock, P.~Shane, The financial analyst forecasting literature: A
taxonomy with suggestions for further research, International Journal of
Forecasting 24, 34--75 (2008).

\bibitem{wom96}
K.~L. Womack, Do brokerage analysts' recommendations have investment value?,
The Journal of Finance 51, 137--167 (1996).

\bibitem{giv79}
D.~Givoly, J.~Lakonishok, The information content of financial analysts'
forecasts of earnings: Some evidence on semi-strong inefficiency, Journal of
Accounting and Economics 1,  165--185 (1979).

\bibitem{lys90}
T.~Lys, S.~Sohn, The association between revisions of financial analysts'
earnings forecasts and security-price changes, Journal of Accounting and
Economics 13, 341--363 (1990).

\bibitem{bra03}
A.~Brav, R.~Lehavy, An empirical analysis of analysts' target prices:
Short-term informativeness and long-term dynamics, The Journal of Finance 58, 1933--1967 (2003).

\bibitem{asq05}
P.~Asquith, M.~B. Mikhail, A.~S. Au, Information content of equity analyst
reports, Journal of Financial Economics 75,  245--282 (2005).

\bibitem{giv09}
D.~Givoly, C.~Hayn, R.~Lehavy, The quality of analysts' cash flow forecasts,
The Accounting Review 84,  1877--1911 (2009).

\bibitem{cal13}
A.~C. Call, S.~Chen, Y.~H. Tong, Are analysts' cash flow forecasts na{\"\i}ve
extensions of their own earnings forecasts?, Contemporary Accounting Research
30,  438--465 (2013).

\bibitem{ert03}
Y.~Ertimur, J.~Livnat, M.~Martikainen, Differential market reactions to revenue
and expense surprises, Review of Accounting Studies 8,  185--211 (2003).

\bibitem{hua14}
A.~H. Huang, A.~Y. Zang, R.~Zheng, Evidence on the information content of text
in analyst reports, The Accounting Review 89,  2151--2180 (2014).

\bibitem{jia22}
X.-F. Jiang, L.~Xiong, T.~Cen, L.~Bai, N.~Zhao, J.~Zhang, C.-J. Zheng, T.-Y.
Jiang, Analyst sentiment and earning forecast bias in financial markets,
Physica A: Statistical Mechanics and its Applications 589, 126601 (2022).

\bibitem{dav12}
A.~K. Davis, J.~M. Piger, L.~M. Sedor, Beyond the numbers: Measuring the
information content of earnings press release language, Contemporary
Accounting Research 29, 845--868 (2012).

\bibitem{fel10}
R.~Feldman, S.~Govindaraj, J.~Livnat, B.~Segal, Management?s tone change,
post earnings announcement drift and accruals, Review of Accounting Studies
15,  915--953 (2010).

\bibitem{lou11}
T.~Loughran, B.~McDonald, When is a liability not a liability? textual
analysis, dictionaries, and 10-ks, The Journal of Finance 66, 35--65 (2011).

\bibitem{twe12}
B.~Twedt, L.~Rees, Reading between the lines: An empirical examination of
qualitative attributes of financial analysts' reports, Journal of
Accounting and Public Policy 31, 1--21 (2012).

\bibitem{bus11}
B.~J. Bushee, M.~J. Jung, G.~S. Miller, Conference presentations and the
disclosure milieu, Journal of Accounting Research 49, 1163--1192 (2011).

\bibitem{lem06}
M.~Lemmon, E.~Portniaguina, {Consumer Confidence and Asset Prices: Some
	Empirical Evidence}, The Review of Financial Studies 19~,
1499--1529 (2006).

\bibitem{qiu04}
L.~Qiu, I.~Welch, Investor sentiment measures, Tech. rep., National Bureau of
Economic Research (2004).

\bibitem{fis00}
K.~L. Fisher, M.~Statman, Investor sentiment and stock returns, Financial
Analysts Journal 56,  16--23 (2000).

\bibitem{bro03}
S.~J. Brown, W.~N. Goetzmann, T.~Hiraki, N.~Shirishi, M.~Watanabe, Investor
sentiment in Japanese and US daily mutual fund flows, Tech. Rep., National
Bureau of Economic Research (2003).

\bibitem{bak04}
M.~Baker, J.~C. Stein, Market liquidity as a sentiment indicator, Journal of
Financial Markets 7,  271--299 (2004).

\bibitem{bak04a}
M.~Baker, J.~Wurgler, A catering theory of dividends, The Journal of Finance 59, 1125--1165 (2004).

\bibitem{bak04b}
M.~Baker, J.~Wurgler, Appearing and disappearing dividends: The link to
catering incentives, Journal of Financial Economics 73,  271--288 (2004).

\bibitem{nea98}
R.~Neal, S.~M. Wheatley, Do measures of investor sentiment predict returns?,
Journal of Financial and Quantitative Analysis 33,  523--547 (1998).

\bibitem{wha00}
R.~E. Whaley, The investor fear gauge, The Journal of Portfolio Management 26, 12--17 (2000).

\bibitem{bak06}
M.~Baker, J.~Wurgler, Investor sentiment and the cross-section of stock
returns, The Journal of Finance 61, 1645--1680 (2006).

\bibitem{bla86}
F.~Black, Noise, The Journal of Finance 41,  528--543 (1986).

\bibitem{chu12}
H.~Chung, J.-P. Laforte, D.~Reifschneider, J.~C. Williams, Have we
underestimated the likelihood and severity of zero lower bound events?,
Journal of Money, Credit and Banking 44, 47--82 (2012).

\bibitem{de90}
J.~B. De~Long, A.~Shleifer, L.~H. Summers, R.~J. Waldmann, Noise trader risk in
financial markets, Journal of Political Economy 98, 703--738 (1990).

\bibitem{meh02}
R.~Mehra, R.~Sah, Mood fluctuations, projection bias, and volatility of equity
prices, Journal of Economic Dynamics and Control 26, 869--887 (2002).

\bibitem{lee02}
W.~Y. Lee, C.~X. Jiang, D.~C. Indro, Stock market volatility, excess returns,
and the role of investor sentiment, Journal of Banking \& Finance 26, 2277--2299 (2002).

\bibitem{men12}
B.~Mendel, A.~Shleifer, Chasing noise, Journal of Financial Economics 104, 303--320 (2012).

\bibitem{che17}
T.-T. Chen, B.~Zheng, Y.~Li, X.-F. Jiang, New approaches in agent-based
modeling of complex financial systems, Front. Phys. 12, 128905 (2017).

\bibitem{che18}
T.-T. Chen, B.~Zheng, Y.~Li, X.-F. Jiang, Information driving force and its
application in agent-based modeling, Physica A 496, 593--601 (2018).

\bibitem{jia13}
{X.F. Jiang, T.T. Chen and B. Zheng}, Time-reversal asymmetry in financial
systems, Physica A 392, 5369 (2013).

\bibitem{zha24}
{J. Zhang,  B.~Zheng, L.-F. Jin, Y.~Li,  X.-F. Jiang}, Non-stationary temporal-spatio correlation analysis of information-driven complex financial dynamics, Chinese Journal of Physics 88, 756--767 (2024).

\bibitem{jin22}
{L.-F. Jin, B.~Zheng, J.-H. Ma, J. Zhang, and L. Xiong, X.-F. Jiang, J.-C. Li}, Empirical study and model simulation of global stock market dynamics during COVID-19,  Chaos, Solitons \& Fractals 159, 112138 (2022).

\bibitem{zha10}
Y.-Y. Zhao, B.~Qin, T.~Liu, Sentiment analysis, Journal of Software 21,
1834--1848 (2010).

\bibitem{kah03}
D.~Kahneman, Maps of bounded rationality: Psychology for behavioral economics,
American Economic Review 93, 1449--1475 (2003).

\bibitem{qiu06}
{T. Qiu, B. Zheng, F. Ren, and S. Trimper}, Return-volatility correlation in
financial dynamics, Phys. Rev. E 73,  065103 (2006).

\bibitem{jia17a}
X.-F. Jiang, B.~Zheng, F.~Ren, T.~Qiu, Localized motion in random matrix
decomposition of complex financial systems, Physica A: Statistical Mechanics
and its Applications 471, 154--161 (2017).

\bibitem{bro04}
G.~W. Brown, M.~T. Cliff, Investor sentiment and the near-term stock market,
Journal of Empirical Finance 11, 1--27 (2004).

\bibitem{bol86}
T.~Bollerslev, Generalized autoregressive conditional heteroskedasticity,
Journal of Econometrics 31, 307--327 (1986).

\bibitem{eng86}
R.~F. Engle, T.~Bollerslev, Modelling the persistence of conditional variances,
Econometric Reviews 5, 1--50 (1986).

\bibitem{bri05}
P.~C. O'BRIEN, M.~F. McNichols, L.~Hsiou-Wei, Analyst impartiality and
investment banking relationships, Journal of Accounting Research 43, 623--650 (2005).

\end{thebibliography}
\end{document}